\documentclass[preprint]{aastex}
\usepackage{psfig}
\newcommand{\lta}{$\; \buildrel < \over \sim \;$}
\newcommand{\simlt}{\lower.5ex\hbox{\lta}}
\newcommand{\gta}{$\; \buildrel > \over \sim \;$}
\newcommand{\simgt}{\lower.5ex\hbox{\gta}}

\newcommand{\kms}{{\rm\,km\,s^{-1}}}

\newcommand{\cm}{{\rm\,cm}}

\newcommand{\ffffff}[1]{\mbox{$#1$}}
\newcommand{\scnd}{\mbox{\ffffff{''}\hskip-0.3em.}}

\newcommand{\apm}{APM~08279+5255}
\newcommand{\hut}{Hutsem\'{e}kers}
\newcommand{\qso}{H~1413+117}

\slugcomment{Accepted to the P.A.S.P.}
\shortauthors{Belle \& Lewis}
\shorttitle{Microlensing of BAL Quasars}

\begin{document}

\title{Microlensing of Broad Absorption Line Quasars: \\
Polarization Variability}

\author{Kunegunda E. Belle}
\affil{Department of Physics and Astronomy, University of Wyoming,
Laramie,
WY 82071; keb@tana.uwyo.edu}
\and
\author{Geraint F. Lewis} 
\affil{ 
Department of Physics and Astronomy, University of Victoria, Victoria, 
B.C., Canada; gfl@uvastro.phys.uvic.ca; and  \\ 
Astronomy Department, University of Washington, Seattle WA, U.S.A.; \\
gfl@astro.washington.edu}

\begin{abstract}
Roughly 10\% of all quasars exhibit Broad Absorption Line (BAL)
features which appear to arise in material outflowing at high velocity
from the active galactic nucleus (AGN). The details of this outflow are,
however, very poorly constrained and the particular nature of the BAL
material is essentially unknown.  Recently, new clues have become
available through polarimetric studies which have found that BAL
troughs are more polarized than the quasar continuum radiation.  To
explain these observations, models where the BAL material outflows
equatorially across the surface of the dusty torus have been
developed. In these models, however, several sources of the BAL
polarization are possible. Here, we demonstrate how
polarimetric monitoring of gravitationally lensed quasars, such as
\qso, during microlensing events can not only distinguish between two
currently popular models, but can also provide further insight into the
structure at the cores of BAL quasars.
\end{abstract}

\keywords{Quasars: Absorption Lines --- Quasars: Individual (H1413+117)}

\newpage

%%
%% Definitions
%%
%%
%%% PLAN
%%
%  Introduction
%  Standard Model
%  Microlensing
%      Case a) Resonance Scattering
%           b) Electron Scattering
%  Timescale and statistics (H1413+117)
%  Conclusion
%%

\section{Introduction}\label{introduction}
While they are amongst the  most luminous objects in the universe, the
vast majority  of the  energy radiated by  quasars arises in  a region
less than a parsec in extent. At cosmological distances, such a region
subtends only microarcseconds and  its structural properties are well
below  detection  by conventional  techniques.   A  growing number  of
studies,  however,   have  demonstrated  that  if   such  sources  are
microlensed by  stars in a foreground system  then observations during
high magnification events can  reveal details at these
scales, probing  both  the  structural  and  kinematic  properties  of
quasars~\citep{ka86,ne88,ir89,sc90,co91,le98}.

Recent  studies have  extended  this earlier  work, providing  further
approaches  by which microlensing  can reveal  the inner  structure of
quasars.   Using the  fact that,  to a  microlensing mass,  the quasar
continuum source appears quite extended, and that variations intrinsic
to  the  source  will  be  manifest in  all  the  macrolensed  images,
\citet{yo98,yo99a} demonstrated  how  the  nature of  the
accretion  disk  can  be  probed with  multi-wavelength  observations.
Accordingly,  the  size of  the  source  may  also be  constrained  by
monitoring    and    studying    quasar    variations    \citep{yo99},
high-magnification  event   shapes  \citep{wy99},  and   fold  caustic
crossing events  \citep{fl99}.  A different  approach by \citet{le98b}
used the microlensing induced centroid shift of a macrolensed image to
determine quasar structure.

In a previous paper, (Lewis \& Belle 1998; hereafter Paper I), we
demonstrated how microlensing may be used to examine the scale of the
clouds responsible for the prominent BALs seen in some quasars.  Here
we extend this earlier study and show how spectropolarimetric
monitoring of BAL quasars can provide further clues to the internal
structure of quasars, specifically the details of the scattering
mechanism responsible for the polarization increase observed within
the BAL troughs.  A description of the current favored models of BAL
systems is presented in Section~\ref{standardmodel}, while the
r\^{o}le of microlensing as a probe of these models is discussed in
Section~\ref{microlensing}.  The statistics of microlensing events are
presented in Section~\ref{timescale}, focusing on the case of the
multiply imaged quasar, \qso. The conclusions of this study are
presented in Section~\ref{conclusions}.

\section{Broad Absorption Line Quasars: Models}\label{standardmodel}
About $10\%$ of quasars display BALs in their
spectra,   appearing  blue-ward  of   prominent  emission   lines  and
exhibiting bulk outflow  velocities of $5000-30000\;{\rm km\;s^{-1}}$.
These are  generally interpreted as resonance  absorption line systems
due  to  highly   ionized  species  \citep{tu84,tu88,tu95,we95}.   The
comparative rarity of BAL quasars has made them the focus of numerous
studies and  has led to the  question of their place  in the unified
model of active galactic nuclei (e.g., Antonucci 1993), a question that
is currently a matter of debate~\citep{kuncic99}.

The  nature of  the absorbing  region responsible  for  these spectral
features, although  the subject of various investigations,  has yet to
be  fully explained.   Several different  observations of  BAL quasars
are, however, beginning to paint a picture of the inner regions of the
quasar  and the material  which produces  the BALs.   Abundance ratios
indicate  that the  constant  source of  material  could emanate  from
atmospheres  of  giant  stars \citep{sc95},  novae  \citep{sh96,sh97},
parts of an inner obscuring torus  or even parts of the accretion disk
\citep{mu95}.  The  high outflow  velocity of the  absorbing material,
most likely  due to radiative  acceleration \citep{ar94}, has  led to
models  which  consider  the   necessary  confinement  of  the  clouds
comprising the BAL region.  Pressure confinement by a hot ambient medium
\citep{ar94} or by  magnetic fields \citep{ar94b}  and  X-ray
shielding of  the material by a  very high column  density ionized gas
\citep{mu95}  have all been  offered as  solutions to  the confinement
problem.  Alternatively, the confinement problem can be avoided
altogether by involving a continuous, high column density outflow of
absorbing gas rather than clouds \citep{kuncic99}.  Electron column
densities imply that the region lies at a distance of 30-500 pc from
the central continuum source \citep{tu86}.  Estimates for the radial
extent of the absorbing clouds range from $10^8$ cm [assuming many
clouds with small filling factors \citep{we95}] up to $10^{13.5}$ cm
[from column density measurements \citep{tu95,mu95}] and even sizes as
large as $10^{15}$ cm [determined from microlensing limits (Paper I)]
have been suggested.

The `blackness' of some of the observed troughs indicates that the
material in the BAL region can completely obscure our view of a quasar
core.  This region typically possesses an angular scale of only
microarcseconds, making a determination of the geometrical properties
of the BAL region virtually impossible with normal observational
techniques.  It has become apparent, however, that our view of quasars
in polarized light is implicitly dependent upon the geometry of the
central regions.  From the various polarimetry studies undertaken for
BAL quasars, several consistent observations have been noted; it is
seen that the broad emission lines evident in the quasar spectra
exhibit no or very low net polarization (e.g., Goodrich \& Miller
1995; Schmidt \& Hines 1999), while it has also been shown that a
greater amount of polarization exists in the BAL troughs ($\sim12\%$)
as compared to the continuum ($\sim2\%$) (e.g., Cohen et al. 1995;
Ogle 1997; Schmidt \& Hines 1999).  Other clues come from the amount
of reddening seen in the broad emission lines, with $E(B-V)\sim0.4$,
which is significantly greater than that seen in the continuum or the
BALs, with $E(B-V)>0.1$ \citep{sp92}.

\begin{figure}
\centerline{
\psfig{figure=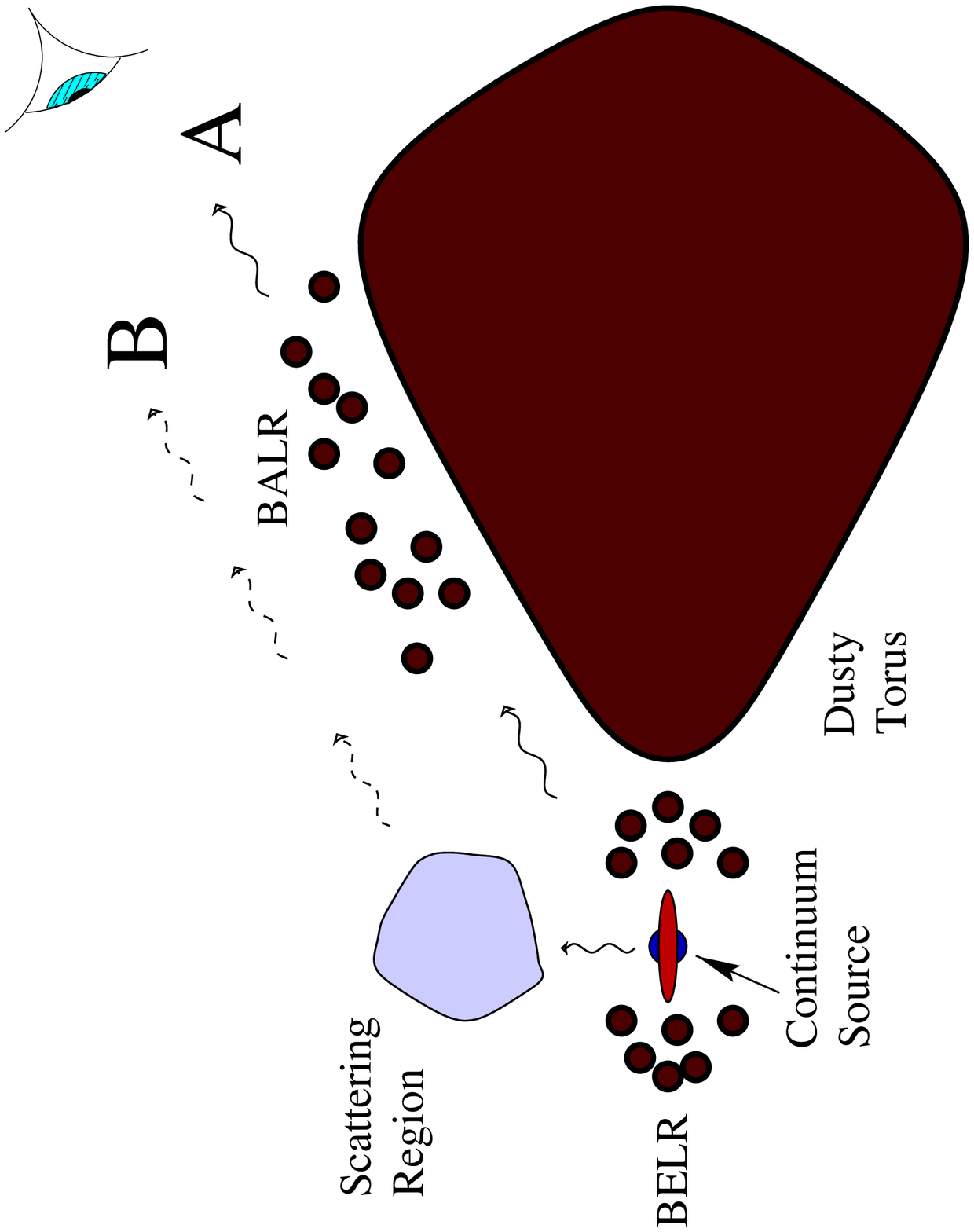,width=4in,angle=270}
}
\caption[]{ A  schematic representation of the  current standard model
for  quasars  exhibiting broad  absorption  features.   Here, the  BAL
material is ablated  from the surface of the  dusty torus. Two sources
for  the  observed  polarization  characteristics are  postulated;  i)
radiation from the continuum  source is intrinsically polarized. While
traveling through  the BAL absorption  region (path {\bf A}),  this is
enhanced by  the action of  resonance scattering of emission  into the
line of sight. ii) the  continuum source is not necessarily polarized,
but  continuum radiation  is polarized  by  impinging on  a region  of
electrons and  dust and being scattered  into the line  of sight (path
{\bf B}). With this, the enhanced polarization seen in the BAL troughs
is  due to  the reduced  unpolarized  flux, while  the polarized  flux
remains relatively free of absorption. \label{fig1}}
\end{figure}

The existence of polarized light in quasar spectra indicates that some
scattering of radiation must be present.  When the above observations
are considered together, a schematic picture of the central regions of
the BAL quasar emerges, with the material responsible for the BAL
absorption being blown equatorially from the dusty torus surrounding
the central accretion disk (e.g., Goodrich \& Miller 1995; Hines \&
Wills 1995; Weymann 1995; Schmidt \& Hines 1999).  As illustrated in
Figure~\ref{fig1}, however, two possible configurations have been
proposed to explain the observed polarization properties of BAL
spectra (e.g., Cohen et al. 1995);
\begin{itemize}
\item{\bf Model A:}  The  path of the continuum
light (Path A) is a simple direct path which leaves the central source
and passes through  the BAL region.  In this case,  it is assumed that
the  continuum emission  is intrinsically  polarized upon  leaving the
source.  As  the radiation travels  through the absorbing  region, any
polarization already  present is enhanced  by the action  of resonance
scattering  of emission into  the line  of sight,  leading to  the BAL
troughs exhibiting greater percentage polarization than the continuum.
\item{\bf Model B:}  This model assumes a  slightly more complex
geometry in which emission  from the continuum source is intrinsically
unpolarized.   This light travels  not  only  directly   to  an  observer,
traversing the BAL region, but also along an alternate path (denoted B
in Figure~\ref{fig1}) impinging upon  a scattering region, most likely
consisting  of  electrons  or  dust  \citep{an85,go95,ga99,br99}.   It  
then travels on to an observer,  leaving the  central  regions of  the  
quasar  without   necessarily  traveling  through  the   BAL   region.   
Unpolarized  light  coming  directly  to  the  observer  dilutes  this
polarized flux  in the continuum of  the spectrum.   The  BAL troughs, 
however, eat into this unpolarized flux,  reducing  the  dilution  and  
enhancing the percentage polarization in the  absorption  line.   
\end{itemize}
The  goal  of  this  paper  is  to  demonstrate  how  photometric  and
polarimetric  monitoring  of   microlensed  BAL  quasars  can  clearly
delineate  between the  above  proposed  models,  and  illustrate  how
microlensing can give further clues to the nature of BAL quasars.

\section{Microlensing}\label{microlensing}
Chang and Refsdal (1979) first demonstrated that the granularity of
galactic matter, distributed on small scales as point-like stars and
planets, can exert a gravitational lensing influence on the light
passing through it from a distant source.  While the image splitting
introduced by such masses is extremely small ($\sim10^{-6}$
arcseconds), such lenses can induce a significant magnification of a
background source.  As the relative configuration of source, lensing
stars, and observer change, so too does the degree of magnification
and consequently the brightness of a background source is seen to
fluctuate.  Such fluctuations have been seen in the light curve of the
quadruple quasar, Q~2237+0305 \citep{ir89,co91}.

The original work of Chang and Refsdal (1979) considered the
gravitational lensing action of a single, isolated star [a description
that accurately represents the simple gravitational lensing by MACHO
objects within our Galactic halo \citep{al93}], however, it is the
case that many stars act on a light beam as it passes through a
galaxy.  In such a high optical depth regime, the pattern of
magnification a source will suffer is no longer simple; rather it is
characterized by a series of violent asymmetric fluctuations,
inter-spaced with quiescent regions where the source suffers
demagnification, also known as the microlensing caustic structure
\citep{ka86}.

Different numerical techniques have  been developed to investigate the
pattern  of magnification  that a microlensed  object  undergoes.  The
backwards   ray-tracing   method,   which   reconstructs  a   map   of
magnification~    \citep{ka86,wa90}   and   the    contour   algorithm
\citep{le93,wi93}  have proven  to be  useful in  the analysis  of the
statistical    properties   of    microlensed    induced   variability
\citep{le95,le96}.

To understand the characteristics of variability introduced by
microlensing, it is necessary to determine two important scale
factors: the Einstein radius and the caustic crossing timescale.  The
Einstein radius, $\eta$, represents the microlensing scalelength in
the source plane. For a single star of mass $M$ it is given as
\begin{equation}\label{e_radius}
\eta = \sqrt{ \frac{4GM}{c^2} \frac{D_{os} D_{ls}}{D_{ol}}},
\end{equation}
where $D_{ij}$ represents the angular diameter distance between
observer ($o$), lens ($l$), and source ($s$) \citep{sc92}.  For
significant magnification, sources must be substantially smaller than
this scalelength.  Considering the cosmological distances involved,
typical values of $\eta$ for multiply imaged quasars, microlensed by
stars of mass $1~M_\odot$, range from $\sim$0.01-0.1 pc.  The
characteristic microlensing timescale, $\tau$, is the time taken by a
region of high magnification (which can be formally infinite at a
`caustic' in the magnification pattern) to sweep across a
source~\citep{ka86}.  This is given as
\begin{equation}\label{time}
\tau \sim \frac{f_{15}}{V_{eff}} \,{\rm yr},
\end{equation}
where $f_{15} \times 10^{15}  h_{75}^{-1}\cm$ is the scalesize of the
source, and
\begin{equation}
V_{eff} =(1+z_l) \frac{D_{os}}{D_{ol}} \frac{v_{300}}{h_{75}} \kms
\label{velocity}
\end{equation}
is the effective velocity of the microlensing caustics, with $z_l$ as
the redshift of the lensing galaxy and $300 v_{300}\kms$ as the
velocity of the microlensing stars across the line of sight, assumed
to be the bulk motion of the galaxy due to its departure from the
Hubble flow. The situation is more complex when one considers motions
due to the stellar velocity dispersion \citep{schramm93,wamby95}, a
point which will be considered later.  For macrolensed quasars,
typical values of $\tau$ are around several months.

\begin{figure}
\centerline{
\psfig{figure=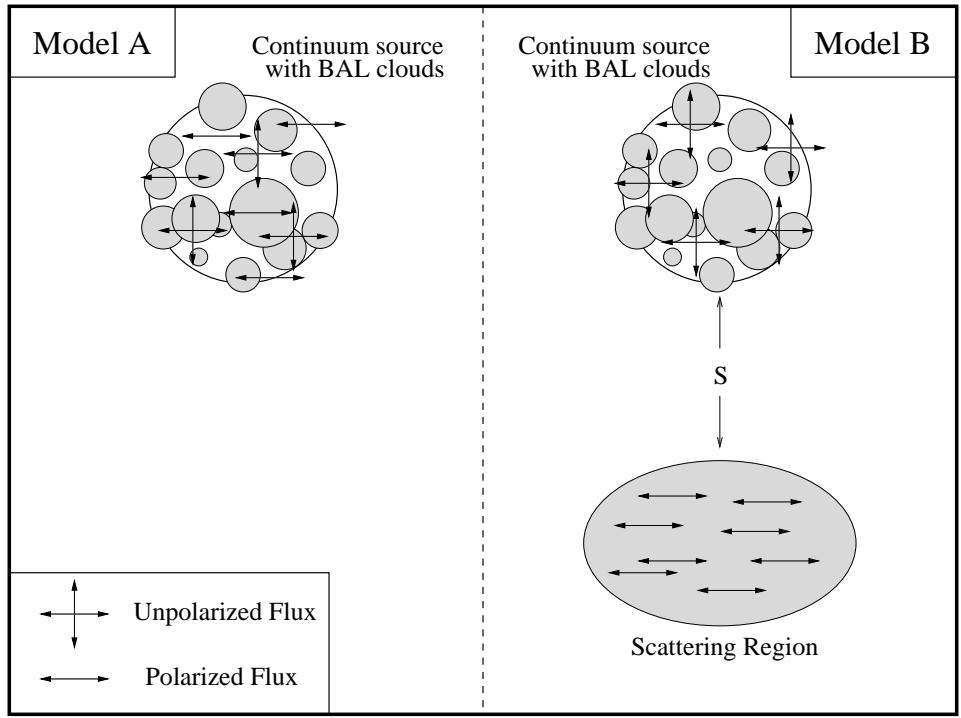,width=5in,angle=270}
}
\caption[]{The view of the  BAL quasar continuum and scattering region
as  seen by the  microlensing caustic  for both  scattering scenarios.
The first  panel shows Model  A, in which resonance  scattering within
the BAL  region causes  the polarization of  the continuum  flux.  The
second  panel  presents  Model  B,  where polarization  is  due  to  a
scattering  region separated  by  a distance,  S,  from the  continuum
source and BAL region. \label{fig2}}
\end{figure}

But  what is  the effect  of microlensing  on each  of  the scattering
models  discussed  in  Section~\ref{standardmodel}?   To  answer  this
question,  we need to  construct, for  each model,  the `view'  of the
central regions of the BAL quasars as seen by the microlensing caustic
structure.  Considering Figure~\ref{fig1},  such a view is represented
in Figure~\ref{fig2},  which will be  discussed in more detail  in the
following  sections.    With  this   picture,  the  question   of  how
microlensing influences our view of  BAL quasars can be addressed. 

\subsection{Model A: Scattering within the BAL Region}\label{resonance}  
For this model, the enhanced polarization seen in the BAL troughs is
resonance scattering within the BAL region itself.  When investigating
the possible microlensing effects, the size of the continuum source as
seen through the scattering BAL region must be considered.  Many
previous studies have demonstrated that this is smaller than the
typical Einstein radius by at least an order of magnitude, and hence,
is subject to extreme gravitational lensing events (e.g., Lewis et
al. 1998).  Similarly, as where the scattering region is the BAL
region, the view of the continuum source through the BAL region in
both polarized and unpolarized light will present very similar
profiles.

This model is illustrated in the left-hand panel of Figure~\ref{fig2}. 
The microlensing masses will see unpolarized and polarized light coming
directly from  the continuum  source   through the  BAL region.
Therefore,  as  the caustic  sweeps  over  the  BAL region,  both  the
scattered light  and absorbed light  are being magnified  equally.  
In this picture,   we  would  expect that  while
microlensing results in a  pronounced photometric change during a high
magnification event,  the observed percentage polarization  in the BAL
trough should  remain unchanged.  Similarly, if a  snap-shot view were
obtained of  a multiply imaged  BAL quasar, each  image would be  in a
different state of microlensing,  although each would possess the same
polarization properties in the BAL troughs.

\subsection{Model B: External Scattering}\label{electron}
In the case  of the external scattering region, we are observing light
traveling two different paths as  it leaves the continuum source, with
the  radiation  which impinges  on  the  scattering  region not  being
significantly attenuated by the  broad absorption material.  As stated
previously,  it is  necessary to  take into  account the  size  of the
scattering region, which in this case  is unknown.  If it is too large
to be affected by microlensing  (i.e., larger in comparison to the 
Einstein  radius of the
microlensing stars),  then the  polarized flux will remain effectively
constant, as the BAL quasar is being microlensed, while the continuum  
flux will  be subject  to variability.
This  leads to  variability  in the  observed percentage  polarization
which will  be seen  in both the  absorption lines and  the continuum.
However,  if the  region is  indeed small enough  to be  significantly
magnified  then the  flux in  polarized  light will  also vary  during
microlensing,  although  there  will   be  a  delay  between  observed
variability in the continuum and in the polarized flux.  If the caustic 
is traveling parallel to the line connecting the continuum and the 
scattering region, the extent of
this delay  depends upon the projected  separation, S, of  the BAL and
scattering regions, and the apparent  velocity of the caustics and may
be defined as,
\begin{equation}\label{delay}
\tau_{delay} \sim \frac{\rm S}{V_{eff}},
\end{equation}
where $V_{eff}$ is the effective velocity (Equation~\ref{velocity}).
To more clearly illustrate this point, the second panel of
Figure~\ref{fig2} depicts the view which the microlensing caustics
will have of the scattering region and the continuum source as seen
through the BAL clouds. It is apparent that the caustic will pass over
the individual regions separately, thereby producing a time delay
between observed variabilities.  Naturally, the value of
$\tau_{delay}$ and which component leads the variability depend on the
direction that the microlensing magnification map sweeps across the
quasar central region.

\begin{figure}
\centerline{
\psfig{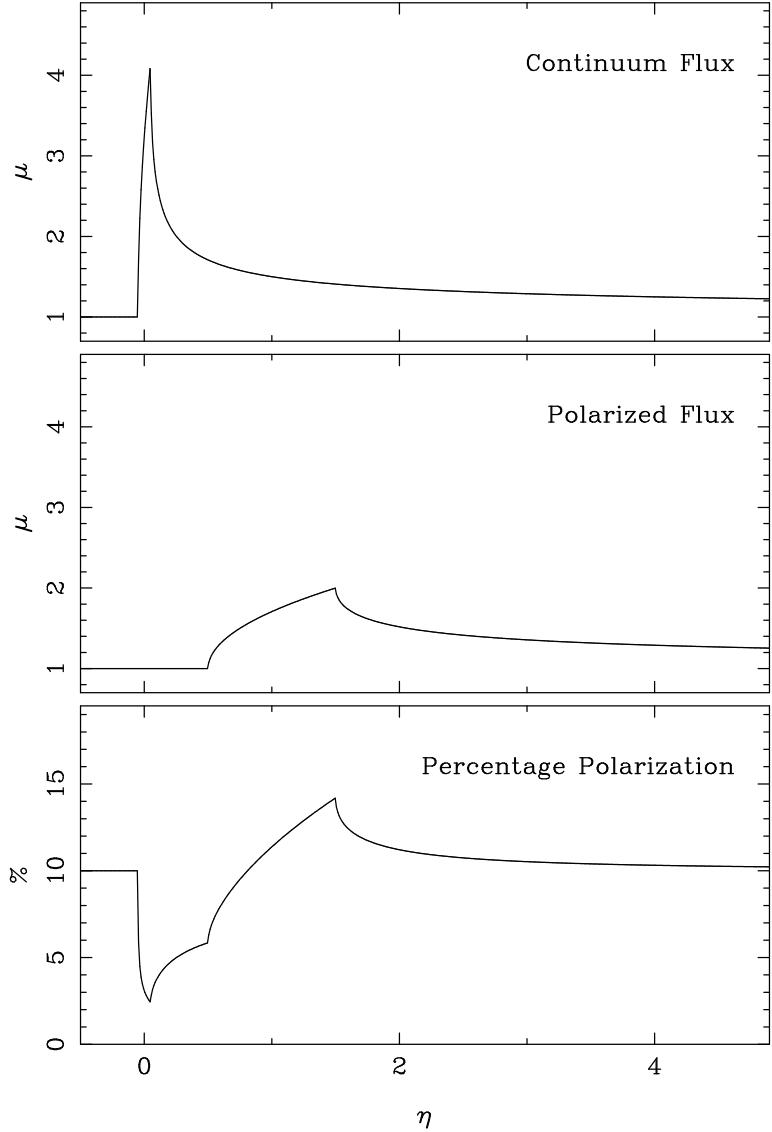}
}
\caption[]{An example of the fluctuation in percentage polarization in
a broad absorption  line trough as a caustic  sweeps across the source
model  B (bottom  panel). Here,  the caustic  sweeps first  across the
small  continuum/BAL source  (top panel),  then over  a  more extended
scattering region which is  responsible for the polarized flux (middle
panel), with  each panel presenting  the magnification of  the source.
The  sharp changes  in both  the  magnification curve,  and hence  the
observed percentage polarization variability, reflect the abrupt edges
in the assumed source model of  a uniform disk. The x-axis is in units
of Einstein radii. \label{fig3}}
\end{figure}

Figure~\ref{fig3} schematically depicts the passing of a caustic
across such a quasar source; both the continuum and scattering region
are treated as being uniform disks with a radius of 0.05$\eta$ and
0.5$\eta$, separated by S$=1\eta$.  The caustic sweeps first across
the continuum source, then across the scattering region resulting in
the magnification fluctuations presented in the upper and middle
panels of Figure~\ref{fig3}.  As pointed out earlier, the scattering
region suffers a lower magnification than the continuum source due to
its more extended nature. The lower panel of Figure~\ref{fig3}
presents the fluctuation in the observed percentage polarization in
the absorption line, assuming an intrinsic 10\% polarization. As the
continuum flux in the line is magnified there is a marked decrease in
the percentage polarization as the emission from the scattering region
is diluted, but later, when the strong magnification of the source has
ceased, the scattering region is slightly magnified, leading to an
enhancement of the percentage polarization. After the caustic has
passed both sources they are equally magnified and the percentage
polarization settles back to its pre-microlensed value, although the
quasar image now appears brighter as it is still magnified by (in this
case) 10\%.

\section{A Microlensed BAL Quasar Case Study}
\subsection{\qso}\label{qso}
To  date,  there  are  several confirmed  gravitationally  lensed  BAL
quasars:    UM    425    \citep{me88,me89,mi95};    \apm\ 
\citep{irwin98,ellison99,ellison99a};       HE       2149-2745
\citep{wi96};    SBS   1520+530    \citep{ch97};    PG   1115+080
\citep{we80,mi96a}; and \qso\ \citep{ma88}.  While the results of this
investigation apply to all of these  subjects, we focus on \qso\ as it
is the most studied and observed gravitationally lensed BAL quasar.

Identified in a survey of luminous quasars, \qso\ was found to possess
four  images  of  a  z=2.55  quasar which  exhibits  broad  absorption
features,   with  image  separations   of  $0\scnd77$   to  $1\scnd36$
\citep{ma88}.   While  spectroscopy has  revealed  the  presence of  a
number of  absorption systems between ${\rm  z\sim1.4-2.1}$, there has
been no  indication of a lensing  system between the  quasar images in
ground based observations  \citep{la96}, although the  system does lie
behind a cluster of galaxies at z$\sim1.7$ which perturbs the lensing via
shearing  \citep{kn98}.  Recent  analysis  of NICMOS  images of  \qso,
however, does  reveal a  very faint ${\rm  (H_{160W}=20.5)}$, extended
system  at the  center of  the cross-like  quasar  image configuration
\citep{kn98a}.

Angonin  et  al.   (1990)  undertook integral  field  spectroscopy  of
individual images  in \qso.   While confirming the  BAL nature  of the
source   quasar,  these   observations  also   revealed  spectroscopic
differences between the images, namely in the relative strength of the
\ion{Si}{4}  ${\rm  \lambda\lambda}$1394,  1403 and  \ion{C}{4}  ${\rm
\lambda}$1549  broad emission  features \citep{hu93},  a  signature of
differential gravitational microlensing by stellar mass objects in the
intervening  galaxy \citep{sa71,le98}.  This  phenomenon has  also been
observed in another quadruple quasar, Q~2237+0305 \citep{le98}.  Such
a conclusion  is supported by  an observed photometric  variability in
\qso\  \citep{re96}  and  significant  differences  between  the  BAL
profiles of the individual images \citep{an90,hu93}.

There have  been several investigations into  the microlensing effects
on BAL  profiles within  the spectra of  the multiple images  of \qso.
The first, which used a Chang-Refsdal Lens \citep{ch79,ch84} to examine the
effects on an individual cloud within the BAL region, was presented by
\hut\ (1993; \hut, Surdej, \& Van Drom 1994).  Although this model was 
able to reproduce the observed spectral variations, it required a very
specific alignment of the caustic, an individual BAL cloud, and the
BAL region, and even a slight change would have a significant effect
on the predicted magnification.  A different approach was taken in
Paper I, where we considered a model for the BAL region consisting of
multiple clouds with varying amounts of absorption.  Although the
caustic framework is rather intricate at high optical depth, the
scalesizes of this and of the source quasar are such that most of the
caustics crossing the BAL region will be isolated fold catastrophes
\citep{sc92}.  This allowed a fairly simple calculation of the
microlensing effects on the BAL region, and it was shown that the
microlensing variability seen within the spectra of the lensed images
of \qso\ could be reproduced if the scalesize of the BAL clouds are
$\sim10^{15}$ cm.
  
\subsection{Event Statistics and Time Scale}\label{timescale}
An important consideration is the timescale on which microlensing
events will occur.  To characterize the lensing scenario, two caustic
crossing timescales need to be defined: the crossing times of the
continuum source and the Einstein radius.  For these calculations, we
assume a standard cosmology of $\Omega=1$ and $\Lambda=0$, and $H_0 =
75 \kms {\rm Mpc^{-1}}$.  Using Equation~\ref{time}, the timescales
can all be calculated rather simply.  If the continuum source is
assumed to be $1\times10^{15}$ cm \citep{wa90b} and the velocity of
the lensing galaxy across the source plane is taken to be $300\kms$,
then the crossing time of the continuum source is $\tau_{cont}\sim0.3$
yr.  The Einstein radius for the parameters pertaining to \qso\ is
$\eta\sim0.006\, \sqrt{M/M_\odot}\,$ pc giving a crossing time of
$\tau_{ER}\sim6.5$ yr for a one solar mass star.  If the separation,
S, between the scattering region and continuum source is taken to be
${\rm S}=1\eta = 19.4\times10^{15}$ cm, then $\tau_{delay} = \tau_{ER}
\sim 6.5$ yr.

As discussed in Section~\ref{microlensing}, the pattern of
magnification that a source undergoes is rather intricate, and the
positions of the caustics projected onto the source plane will
obviously dictate when and how often a source will undergo a
microlensing event.  In order to determine the timescales on which
these events will occur, it is necessary to examine the light curve of
a source which results as the caustic network passes over the source.
Such calculations were performed for the lensed quasar Q~2237+0305 by
\citet{wi93a}, who determined the nature of the light curve specific
to the lensing parameters of each individual image.  From these light
curves the statistics of the microlensing events could be calculated,
including the average time between high magnification events.  To
determine similar parameters for \qso, we can extrapolate from the
work of \citet{wi93a} (because of similarities in lensing parameters)
and then scale these numbers to \qso.  We find that if the source is
moving parallel to the shear, then a lensing event will occur every
$\sim1.3$ yr and if the source is moving perpendicular to the shear,
then lensing events will occur on timescales of $\sim0.9$ yr.  These
timescales mean that {\em significant microlensing effects can be seen
if observations are made every couple of weeks over the course of a
few years.}

\section{Conclusions}\label{conclusions}

While current polarimetric data of BAL quasars have led to theories
about the origin of the polarized light, these observations cannot
clearly delineate between the various scattering models.  However, as
we have argued in this paper, microlensing may provide more detailed
clues about the structure of the inner regions of quasars.
Considering a scattering region associated with the BAL material and a
scattering region separate from the BAL region as the most widely
accepted scattering models, there will result three
different signatures apparent in observations of a microlensed BAL
quasar:
\begin{itemize}
\item Case 1 (Model A):  There is no time delay between continuum and
polarization variability and no percentage polarization change in the
BALs.
\item Case 2 (Model B):  No time delay is observed between continuum and 
polarization variability but a percentage polarization change in the
BALs is detected.
\item Case 3 (Model B):  There is a time delay between continuum and 
polarization variability, and a percentage polarization change is
apparent in the BAL troughs.
\end{itemize}
(Note that as there is an overall enhancement in polarized flux, the
percentage polarization of the continuum will also be seen to change
during a microlensing event for all of these cases.)  Case 1 implies
that the polarization is actually associated with the absorbing
material.  Case 2 and 3 both suggest that the scattering region is not
associated with the absorbing material, although, Case 2 indicates
that the scattering region is large, while Case 3 points to a small
scattering region.  If Case 2 or 3 is apparent in observations of a
gravitationally lensed BAL quasar, then the separation between the
scattering region and the continuum source may be calculated from the
observed time delay.  Most importantly, observations of a single
microlensing event will allow us to differentiate between the two
scattering models.

In considering these models, however, it is important to also
examine the effects of intrinsic variability of the AGN source which
may mask any microlensing signal.  Foremost is luminosity
variability.  Important for model B, any change in the source
luminosity will manifest itself in polarized flux after a delay due to
the geometric light travel distance.  Such variability, however,
possesses several properties that make it easily distinguishable from
any microlensing induced features.   Namely, that change in unpolarized
flux always leads that of the polarized flux and the observed delay
between the two will be the same in all images of a multiply lensed
quasar.  With microlensing, however, the relative phase and time delay
between the unpolarized and polarized variability depend upon the
direction and velocity that a caustic sweeps across the AGN source,
properties that will be different in each macrolensed
image.  Similarly, an observed polarization fluctuation that may be
attributed to microlensing of model A could possibly be due to
intrinsic variability of the polarized emission from the quasar
source.  As with luminosity variability, however, such intrinsic
polarization variability will manifest itself in each component of a
multiply imaged quasar, modulo a gravitational lensing time delay, and
hence can be accounted for.  The details of microlensing induced
variability, however, will be unique to a particular image and will be
uncorrelated with that of the other images.

Although we have presented a simple and straightforward method for
using polarimetric observations of gravitationally microlensed BAL
quasars to determine various parameters pertaining to the
polarization, there is a degeneracy in these calculations as all of
the quantities depend on the mass of the microlensing objects (as
$\sqrt{M}$) and the unknown transverse velocity.  This is
also complicated by the fact that there may be a significant velocity
dispersion within the lensing galaxy which would make the calculations
outlined in this paper much more complex.  Removing these degeneracies
in another gravitationally lensed quasar, Q~2237+0305, has been the
goal of recent studies by \citet{wy99d} and \citet{wy99a,wy99b,wy99c}
who have demonstrated that monitoring gravitationally lensed quasars
will lead to the determination of these unknown quantities.  We plan
to extend the current study presented in this paper with simulations
that will more closely examine microlensing effects occurring within
the multiple images of \qso, and which will also consider the mass and
velocity degeneracies.

 \acknowledgments
Zdenka Kuncic and Nahum Arav are thanked for extremely useful discussions.

\newpage

\end{document}